\documentclass[journal,12pt,onecolumn,draftclsnofoot,]{IEEEtran}
\usepackage{amsmath,amssymb,amsfonts}
\usepackage{algorithmic}
\usepackage{array}
\usepackage[caption=false,font=normalsize,labelfont=sf,textfont=sf]{subfig}
\usepackage{textcomp}
\usepackage{stfloats}
\usepackage{url}
\usepackage{verbatim}
\usepackage{graphicx}
\usepackage{cite}
\usepackage{hyperref}
\hypersetup{
    colorlinks=true,
    linkcolor=blue,
    filecolor=magenta,      
    urlcolor=cyan,
    pdftitle={Overleaf Example},
    pdfpagemode=FullScreen,
    }

\usepackage{multirow}
\usepackage{booktabs}
\usepackage[ruled,vlined,linesnumbered]{algorithm2e}
\SetKwInput{KwInput}{Input}                
\SetKwInput{KwOutput}{Output}              
\SetKwComment{Comment}{$\triangleright$\ }{}
\usepackage{xcolor, pifont}

\begin{document}

\title{Gradient Compression and Correlation Driven Federated Learning for Wireless Traffic Prediction}

\author{Chuanting~Zhang,~\IEEEmembership{Senior Member,~IEEE,}
        Haixia Zhang,~\IEEEmembership{Senior Member,~IEEE,}
        Shuping~Dang,~\IEEEmembership{Senior Member,~IEEE,}
        Basem Shihada,~\IEEEmembership{Senior Member,~IEEE,}
        and~Mohamed-Slim Alouini,~\IEEEmembership{Fellow,~IEEE}%
\thanks{This work was supported in part by NSFC under No. 62401338 and No. 61971270, in part by the Joint Funds of the NSFC under Grant No. U22A2003, in part by the Shandong Province Excellent Youth Science Fund Project (Overseas) under Grant No. 2024HWYQ-028, and by The Fundamental Research Funds of Shandong University. (\it{Corresponding author: Haixia Zhang})}
\thanks{C.~Zhang is with School of Software, Shandong University, 
Jinan 250101, China. Email: chuanting.zhang@sdu.edu.cn}
\thanks{H. Zhang is with School of Control Science and Engineering, Shandong University, Jinan 250061, China. H. Zhang is also with the Shandong Key Laboratory of Wireless Communication Technologies, Shandong University, Jinan 250061, China. E-mail: haixia.zhang@sdu.edu.cn.}
\thanks{S. Dang is with Department of Electrical and Electronic Engineering, University of Bristol, Bristol BS8 1UB, UK. E-mail: shuping.dang@bristol.ac.uk.}
\thanks{B. Shihada and M.-S. Alouini are with Computer, Electrical and Mathematical Science and Engineering Division, King Abdullah University of Science and Technology (KAUST), 
Thuwal 23955-6900, Saudi Arabia. E-mail: \{basem.shihada, slim.alouini\}@kaust.edu.sa}
\thanks{Manuscript received April 19, 2021; revised August 16, 2021.}
}

\markboth{Journal of \LaTeX\ Class Files,~Vol.~14, No.~8, August~2021}%
{Shell \MakeLowercase{\textit{et al.}}: A Sample Article Using IEEEtran.cls for IEEE Journals}


\maketitle

\begin{abstract}
Wireless traffic prediction plays an indispensable role in cellular networks to achieve proactive adaptation for communication systems. Along this line, Federated Learning (FL)-based wireless traffic prediction at the edge attracts enormous attention because of the exemption from raw data transmission and enhanced privacy protection. However FL-based wireless traffic prediction methods still rely on heavy data transmissions between local clients and the server for local model updates. Besides, how to model the spatial dependencies of local clients under the framework of FL remains uncertain. To tackle this, we propose an innovative FL algorithm that employs gradient compression and correlation-driven techniques, effectively minimizing data transmission load while preserving prediction accuracy. Our approach begins with the introduction of gradient sparsification in wireless traffic prediction, allowing for significant data compression during model training. We then implement error feedback and gradient tracking methods to mitigate any performance degradation resulting from this compression. Moreover, we develop three tailored model aggregation strategies anchored in gradient correlation, enabling the capture of spatial dependencies across diverse clients. Experiments have been done with two real-world datasets and the results demonstrate that by capturing the spatio-temporal characteristics and correlation among local clients, the proposed algorithm outperforms the state-of-the-art algorithms and can increase the communication efficiency by up to two orders of magnitude without losing prediction accuracy. Code is available at \url{https://github.com/chuanting/FedGCC}.
\end{abstract}


\section{Introduction}\label{sec:introduction}

\IEEEPARstart{E}{dge} intelligence, empowered by artificial intelligence techniques (e.g., machine learning and deep learning), is deemed one of the essential missing functionalities in the current fifth-generation (5G) networks \cite{Quraan2023,peltonen20206g,Letaief2019}. Consequently, research communities from both academia and industry have reached a consensus that edge intelligence will play notable roles in the future sixth-generation (6G) communication networks \cite{dang2020should,Niknam2020,ye2021jointran,yuan2024low,yang2024detfed}. By pushing part of the service-related processing and data storage from the central cloud to edge nodes that are geographically and logically close to the end users, the communication efficiency, computational efficiency, and end-to-end latency of the core network can be drastically improved.

\begin{figure}[!t]
\centering
\subfloat[]
{\includegraphics[width=0.25\textwidth]{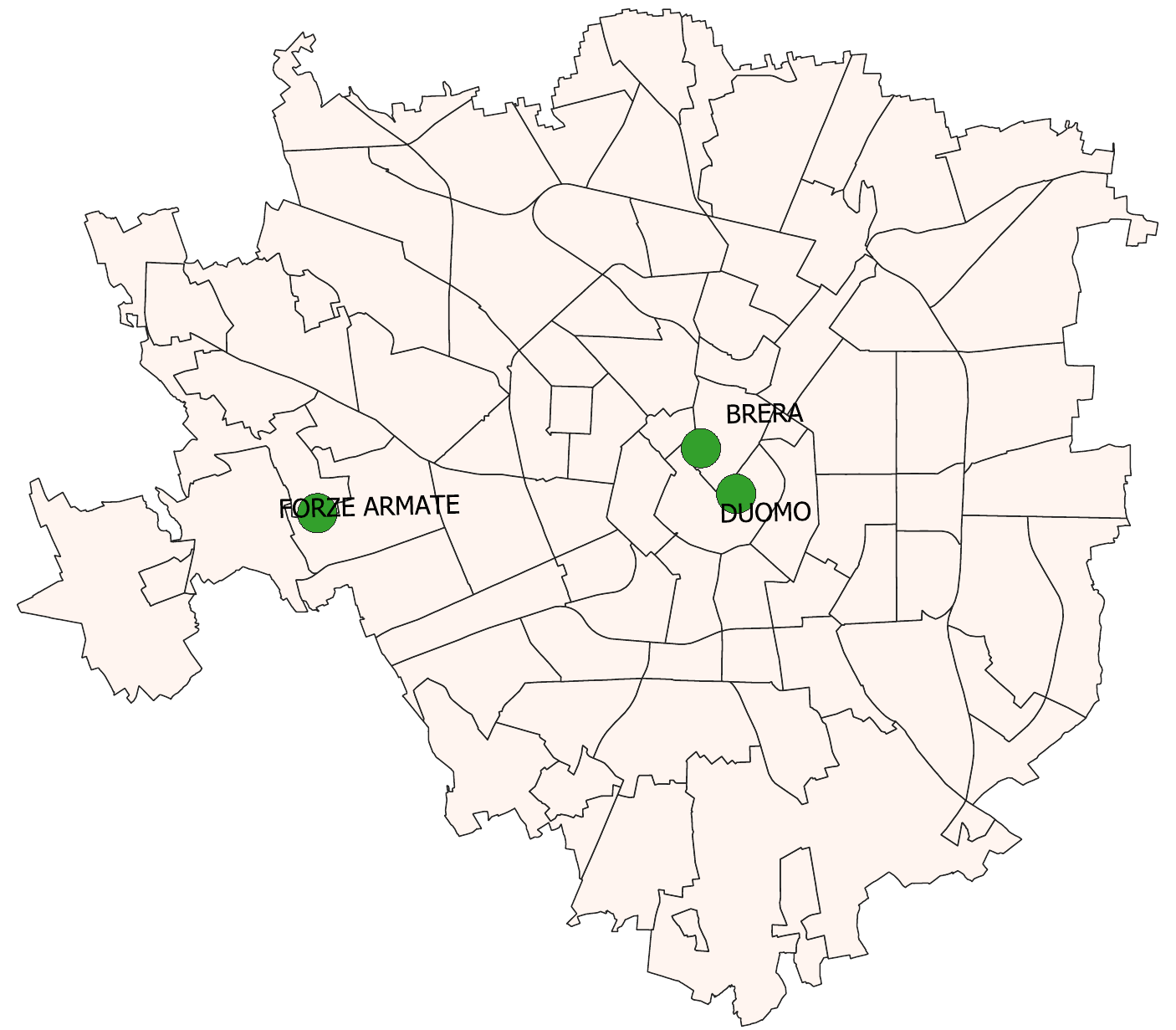}
\label{milano_map}}
\subfloat[]
{\includegraphics[width=0.25\textwidth]{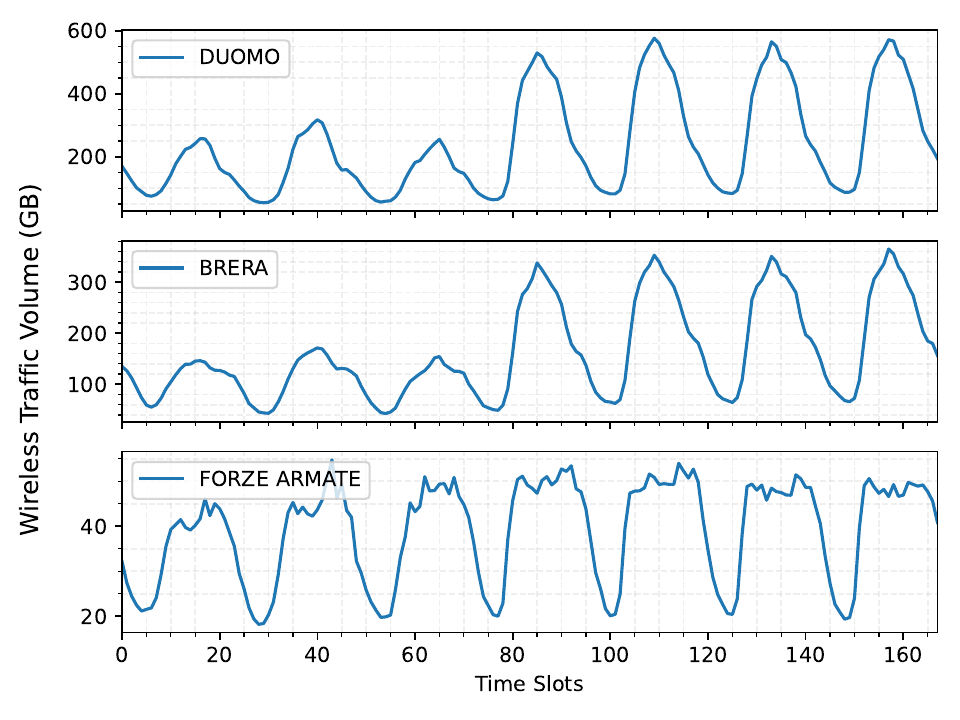}
\label{temporal_dynamics}
}

\subfloat[]
{\includegraphics[width=0.25\textwidth]{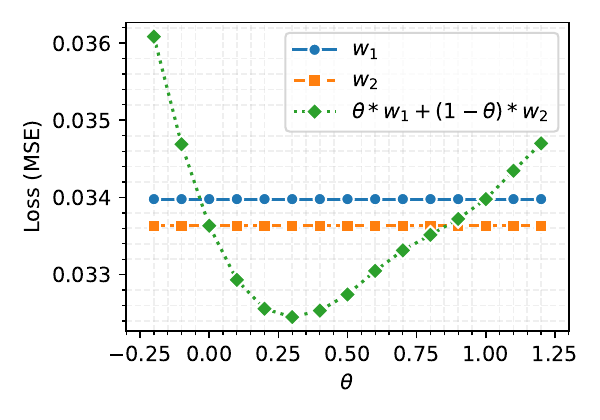}
\label{homo}
}
\subfloat[]
{\includegraphics[width=0.25\textwidth]{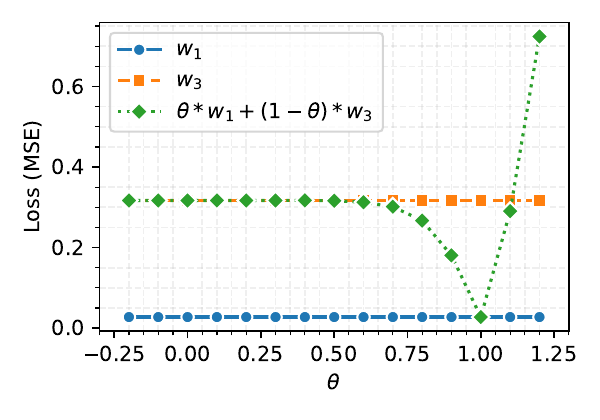}
\label{heter}
}
\caption{An example on model aggregation using FedAvg algorithm: (a) Milan city boundary and three selected places; (b) Temporal dynamics of the three selected places; (c) Aggregate models trained with similar traffic patterns improves performance i.e., mean squared error (MSE) achieved by $\theta w_1 + (1-\theta) w_2$ is lower than that by either $w_1$ or $w_2$; (d) Aggregate models trained with distinct traffic patterns brings no performance improvements.}
\label{motivated_example}
\end{figure}

Along this line, wireless traffic prediction \cite{habib2024transformer,li2024core,Xu2019, zhang2022fedmeta} at the edge is critical in supporting the realization of edge intelligence, mainly because many intelligent network operations such as self-organization and self-optimization heavily depend on future traffic status. Though many works on wireless traffic prediction have been conducted \cite{Xu2016,Zhang2018,Feng2018,Zhang2019,Zhao2020,Wang2019,Wang2017,Qiu2018,Li2017,Li2014,Shen2021}, almost all of them fall into the realm of centralized model training, which requires raw data transmission from edge nodes (local client) to the central cloud. These edge nodes are located in physically different places, and raw data transmission not only consumes valuable bandwidth but also may incur privacy issues if attacked by malicious agents.

Thanks to the advancement of distributed machine learning, particularly federated learning (FL) \cite{li2019federated,Tran2019a,Haddadpour2021,zhao2018federated,Li2020,stich2018sparsified,liu2020traffic,gao2024federated,gao2024communication,li2023wireless}, both model training and prediction can be performed at the edge. Under the FL paradigm, multiple edge nodes train a global model collaboratively under the orchestration of the central cloud. Each edge node needs only to transfer the updated model or gradient information instead of raw data to the central cloud. 

Despite its promising advantages in model training, applying FL to wireless traffic prediction still faces many challenges. First, although no raw traffic data transmission is involved in the model training, the frequent exchange of gradients or parameters between edge nodes and the central server brings nontrivial communication overhead to the wireless network, especially when the model is large. Second, edge nodes have different traffic patterns and complex spatial-temporal dependencies induced by user mobility. Traditional FL algorithms may not work, as Fig. \ref{motivated_example} shows, since they simply average the gradients of different edge nodes without distinguishing their different contributions to the final global model. More specifically, three edge nodes are selected from our Milan dataset and their geographical locations and traffic patterns are included. When two edge nodes have similar traffic patterns, averaging their models improves prediction performance. However, if their patterns differ, model aggregation may not yield any benefits and could even degrade performance.

To solve the above two challenges faced with wireless traffic prediction,
we design a novel federated learning algorithm based on gradient compression and correlation. 
Specifically, we introduce gradient sparsification \cite{lin2017deep} into wireless traffic prediction to release the communication overhead between local clients and the central server. As its lossy nature of sparsification, we further propose using error feedback and gradient tracking techniques \cite{Xin2019,Lu2019,richtarik2021ef21,Haddadpour2021} to compensate for the negative effects of gradient compression. Moreover, we propose three personalized model aggregation strategies derived from the gradient correlation matrix to capture distinct spatial dependencies of local clients. To the best of the authors' knowledge, this is the first work presenting such a distributed architecture and the related methods incorporating spatio-temporal characteristics for wireless traffic prediction. 
The experimental results demonstrate that the proposed algorithm equipped with gradient compression and correlation outperforms the state-of-the-art and can raise the communication efficiency up to two orders of magnitude without loss of prediction accuracy.

The rest of this paper is organized as follows. We survey related works in Section II and introduce the system model and problem formulation in Section III. Our proposed algorithm will be detailed in Section IV. After that, we empirically validate the performance of our proposed algorithm in Section V and conclude the paper in Section VI.

\section{Related Works}
Wireless traffic prediction has achieved considerable attention in recent years due to its critical role in advancing the intelligence of future communication networks, such as intelligent network management and optimization.
Earlier works on wireless traffic prediction mainly focus on traditional statistical models such as ARIMA \cite{zhou_ngide_2006} and its variant \cite{Shu2005}, $\alpha$-stable \cite{Li2017}, and entropy \cite{Li2014}. However, the performances of these models are limited by their representation ability to capture the nonlinear relationships of wireless traffic.

Thanks to the rapid advancement of artificial intelligence techniques, notably machine learning and deep learning, they have extensively expedited the development of wireless traffic prediction. Many machine or deep learning based methods have sprung out in the past several years. Methods based on Bayesian process \cite{Xu2016,Xu2019}, deep belief networks \cite{Nie2017}, long-short term networks \cite{Qiu2018}, and graph neural networks \cite{Wang2019} are all explored to solve wireless traffic prediction problems. 
In addition, the encoder-decoder architecture can also be used to capture the spatial-temporal dependencies in wireless traffic prediction \cite{Wang2017,Feng2018}.

For network-wide or city-scale wireless traffic prediction,  The authors in \cite{Zhang2018} proposed using convolutional neural networks to capture the spatial and temporal dependencies simultaneously. They also considered different kinds of temporal dependencies such as closeness and periodicity. \cite{Zhang2018a} adopted a three-dimensional convolutional network to enhance spatial-temporal learning over wireless traffic. Besides, in \cite{Zhang2019}, cross-domain data, such as points of interest, have proven to be helpful in improving prediction accuracy. Later, the attention schemes are introduced into wireless traffic prediction \cite{Zhao2020,Shen2021} to reduce model complexity and improve representation learning.

The above works mainly focus on wireless traffic prediction in a centralized way, which consumes lots of bandwidth since raw wireless traffic data must be transferred to a powerful node for training. FL-based methods are starting to emerge to solve this problem and train a prediction model at the edge. \cite{Liu2020} proposed a federated gated recurrent neural network for traffic flow prediction. In addition, to solve the data heterogeneity challenge confronted with FL, the authors in \cite{zhang2021fedda} designed a dual-attention scheme named FedDA, which makes the learning process aware of both local updates and prior global knowledge. FedDA is designed particularly for wireless traffic prediction and achieves state-of-the-art performance.

Our work also considers training a wireless traffic prediction model under the FL paradigm, but it differs from all the above works in two major aspects. First, we introduce gradient compression into wireless traffic prediction to release the communication overhead between local clients and the central server. Second, we design personalized gradient aggregation schemes derived from gradient correlation to capture the spatial dependencies among different local clients.

\section{System Model and Problem Formulation}

\begin{figure*}[!t]
\centering
\includegraphics[width=1.0\textwidth]{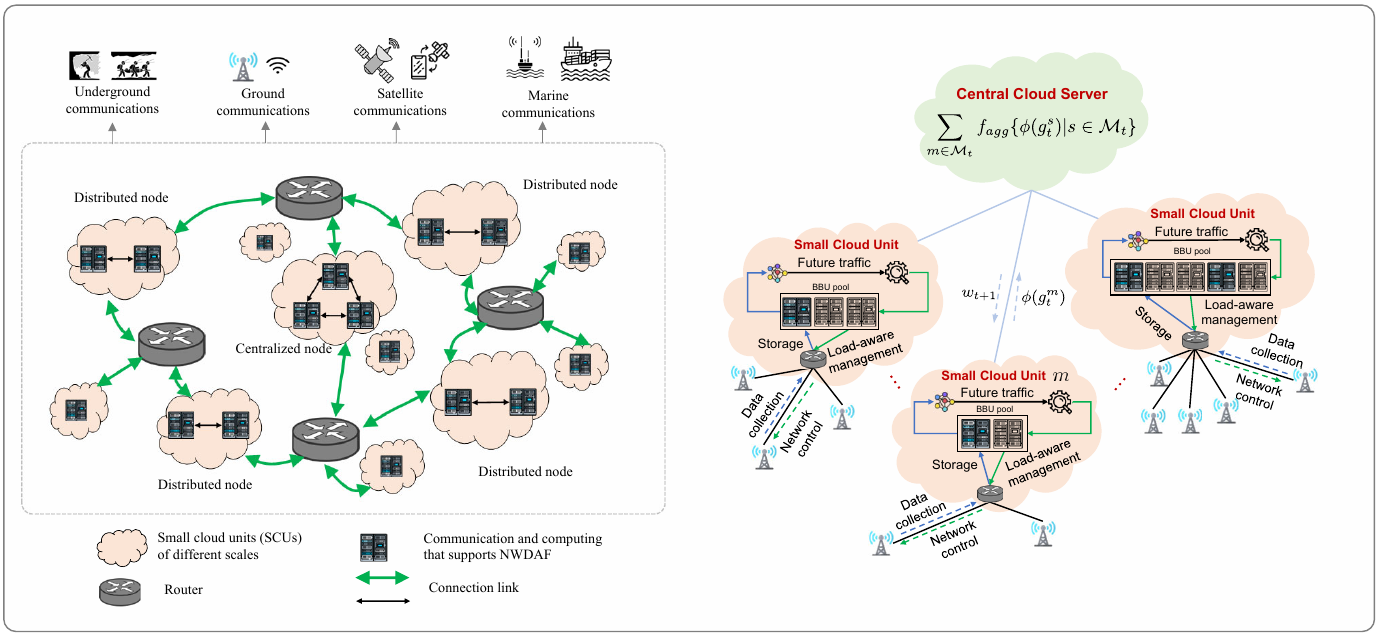}
\caption{Left: The architecture of distributed autonomous networks (DAN); Right: An abstract and simplified DAN for wireless traffic prediction. DAN is a promising 6G networking architecture that supports intelligent network elements such as network data analytics function (NWDAF). In our wireless traffic prediction system, $M$ SCUs train a robust prediction model collaboratively under the orchestration of a central cloud server in a communication-efficient way.}
\label{sys_diagram}
\end{figure*}

We perform the study of wireless traffic prediction under the scenario of distributed autonomous networks (DAN), which is a promising network architecture for six-generation (6G) wireless communications. The concept of DAN is introduced on the left of Fig. \ref{sys_diagram}, based on which we abstract a simplified system model for wireless traffic prediction and show it on the right of Fig. \ref{sys_diagram}. The DAN architecture is designed to support on-demand network customization, plug-and-play, and flexible deployment through distributed and autonomous features to meet the diverse needs of future 6G networks. DAN has many distributed small cloud units (SCU) deployed in different places to achieve underground communications, ground communications, satellite communications, and marine communications. Each SCU has both communication and computing capabilities such as a baseband unit (BBU) pool and smart edge engine, which makes them smart enough to collect and analyze data by supporting the network data analytics function (NWDAF). 
Our wireless traffic prediction system model consists of $M$ small cloud units (SCUs) at the edge and one central cloud server at the mobile core network.
Based on different coverage demands, each SCU supports a different number of base stations (BSs). The SCU monitors and collects mobile traffic data from each BS in real-time and stores them in its local database. 
With these datasets, the SCUs train a prediction model $w$ in a federated way to predict the future traffic volume whilst performing load-aware traffic management and network control. For instance, some of the BBUs can be switched to an idle state to save power when the predicted traffic volume is small.

Given the $M$ SCUs in the communication system, each SCU $m$ stores a wireless traffic dataset $\mathcal{D}^m=\{x^m_i, y^m_i\}_{i=1}^n$, where $x^m_i=[v^m_i,v^m_{i+1},\cdots,v^m_{i+p-1}]$ and $y^m_i=v^m_{i+p}$. Specifically, $v^m_i$ denotes the wireless traffic volume at time slot $i$ of SCU $m$ and $p$ is the selected sliding window size when constructing training and test data samples.

We consider leveraging FL, or more precisely, cross-silo FL to obtain the prediction model $w$. 
Our objective function can be written as follows
\begin{equation}\label{problem}
\arg\min_{w\in \mathbb{R}^d} \Big\{ f(w) = \frac{1}{M}\sum_{m=1}^M f_m (w)\Big\},
\end{equation}
where $d$ denotes the dimension of model parameter and $f_m(\cdot)$ the local objective function of client $m$. $f_m(\cdot)$ generally takes the form of $\ell (x^m_i, y^m_i; w)$, in which $\ell(\cdot)$ measures the distance between predictions and the ground truth value.

There are two fundamental steps to solve (\ref{problem}) in the traditional FL paradigm: local updates on the client and global aggregation on the server side. Specifically, at the $t$-th communication round, the first step refers to local client $m$ computes the gradient information $g^m_t = \nabla f_m(w_t)$ and sends it to the server. The second step refers to the server updating the global model based on the received gradients from local clients. The classical weight update of FL algorithms, i.e., FedAvg \cite{McMahan2017}, can be written as
\begin{equation}\label{server}
w_{t+1} = w_t - \frac{\eta}{|\mathcal{M}_t|}\sum_{m\in \mathcal{M}_t} g^m_t,
\end{equation}
where $\eta$ is the learning rate at the server and $\mathcal{M}_t$ is a set of local clients at communication round $t$. Besides, $|\cdot|$ represents the cardinality of a set.

\section{Our Proposed Method}
This section introduces our proposed gradient compression and correlation driven FL algorithm for wireless traffic prediction problems. The overall idea will be given first, then the detailed designs of each component will be introduced.

\subsection{Key Insights}
Different from (\ref{server}), in this work, we try to solve (\ref{problem}) in a robust and communication-efficient approach by re-written (\ref{server}) into the following
\begin{equation}\label{new_problem}
w_{t+1} = w_t - \frac{\eta}{|\mathcal{M}_t|}\sum_{m\in \mathcal{M}_t} f_{agg}\lbrace \phi(g^{s}_t)| s \in \mathcal{M}_t \rbrace,
\end{equation}
where $\phi (\cdot)$ represents a compressor on the gradient, whose purpose is to reduce the communication overhead during model training. In addition, $f_{agg}(\cdot)$ represents a family of personalized aggregation strategies derived from gradient correlation, aiming to capture the peculiar spatial dependencies between client $m$ and all the others. We elaborate the detailed designs of $\phi (\cdot)$ and $f_{agg}(\cdot)$ in the following.

\subsection{Local Update on the Client}
Gradient compression techniques are proposed to reduce the communication overhead in distributed machine learning \cite{lin2017deep}, but compression negatively influences prediction performance, especially when the compression ratio $\gamma$ is large \cite{Haddadpour2021}. This is because a majority of local gradient information is filtered out during communications between local clients and the central server, resulting in a limited number of local gradients involved in the global model update. The scenario is even worse when introducing gradient compression for wireless traffic prediction problems since geographically distributed SCUs have diverse traffic patterns. Relying only on the local gradient information to update the local model could deviate the global model from the optimum since the local gradient directions could be highly different from the global gradient direction.
To solve the above issue, we resort to error feedback \cite{richtarik2021ef21} and gradient tracking techniques \cite{Haddadpour2021}.
On the one hand, the error feedback schemes make transferring all local gradients to the server possible without increasing communications. On the other hand, gradient tracking techniques enhance local model training by monitoring the difference between local gradient directions and the global gradient direction.

\begin{figure}[!t]
\centering
\includegraphics[width=0.6\textwidth]{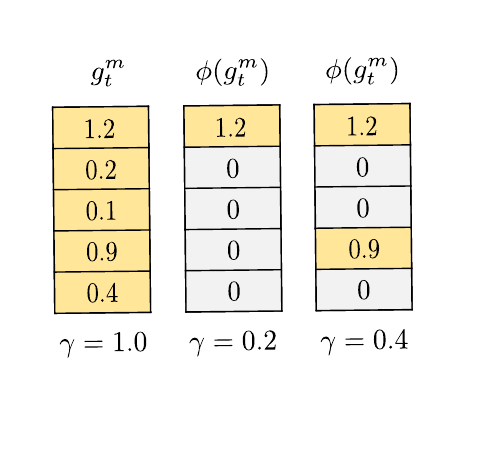}
\caption{A demonstration of gradient sparsification. Original gradient vector $g_t^m$ with 5 elements (left) and the corresponding compressed version when $\gamma=0.2$ (middle) and $ \gamma=0.4$ (right), respectively. Note that shadowed elements are those that will be transferred and $\gamma=1.0$ indicates no compression.}
\label{phi}
\end{figure}

To explicitly describe our algorithm, consider the $c$-th local update in the $t$-th communication round at the client $m$. We first sample a batch of data samples $\mathcal{B}^m_{t,c}$ with the size of $b$. Then we obtain the corrected gradient information for this local update, which is described by
\begin{equation}\label{corrected_gradient}
g_{t, c}^m = e_t^m + \nabla f_m(w_{t,c}^m;\mathcal{B}_{t,c}^m) - h_t^m.
\end{equation}
In (\ref{corrected_gradient}), $e_t^m$ denotes an error feedback vector at client $m$ in round $t$ and represents the sum of the gradient information that has been filtered out so far; $\nabla f_m(w_{t,c}^m;\mathcal{B}_{t,c}^m)$ is the gradient on the current batch data samples; $h_t^m$ contains the sum of the difference between local client's gradients and the globally averaged gradients. $e_t^m$ and $h_t^m$ keep static during local updates but vary over communication rounds.

Once we obtain $g_{t, c}^m$ we can update the local model with local learning rate $\epsilon$ as follows
\begin{equation}
w_{t,c+1}^m = w_{t,c}^m - \epsilon g_{t, c}^m.
\end{equation}
The final output after $\tau$ steps of local updates is expressed as $w_{t, \tau}^m$. Having obtained this, we can calculate the accumulated gradients during these $\tau$ steps, which is $g_t^m = (w_t^m - w_{t,\tau}^m)/\epsilon$. 
After that, the local client sends a compressed version of $g_t^m$, i.e., $\phi (g_t^m)$, to the central server for global model optimization.

As for the form of $\phi (\cdot)$, there are normally two options, i.e., sparsification and quantization. Here, in this work, we consider the first option: we use gradient sparsification as our compressor $\phi (\cdot)$. This is because sparsification possesses more potential in gradient compression since it is not limited by the integer representation of computer systems \cite{lin2017deep}. More concretely, when the compression ratio is set to $\gamma$, then we only transmit the top $\gamma$ of the gradient elements in terms of their modulus. A demonstration of $\phi (\cdot)$ is shown in Fig. \ref{phi}.

The local clients will update the error feedback vector $e_t^m$ and the gradient tracking vector $h_t^m$ respectively after finishing the gradient compression. The updates are given by
\begin{equation}
e_{t+1}^m = e_{t}^m + g_t^m - \phi (g_t^m),
\end{equation}
\begin{equation}\label{gt_update}
h_{t+1}^m = h_{t}^m + \frac{1}{\tau}(\phi(g_t^m) - g_t),
\end{equation}
where $g_t$ denotes the averaged gradient information at communication round $t$, which is calculated as $\frac{1}{|\mathcal{M}_t|}\sum_{m\in \mathcal{M}_t}\tilde{\phi}(g_t^m)$. 
Note that, in Eq. (\ref{gt_update}), the central server also needs to broadcast $g_t$ to each local client to update the local gradient tracking vector. Besides, calculating $e$ and $h$ will bring a little bit of complexity, but it can solve the challenges of data heterogeneity and gradient compression.
The details to obtain $\tilde{\phi}(g_t^m)$ will be explained in the next subsection.

\subsection{Personalized Aggregation Strategies with Gradient Correlation}
When we receive all the local gradient information at the central server, we can proceed to the global model optimization.
As a previous study \cite{Ji2019} reveals that simply averaging the local gradient information without distinguishing their different contributions to the final global model yields poor predictions, especially for wireless traffic prediction problems \cite{zhang2021fedda}. 
Thus in this subsection, we propose three personalized gradient aggregation strategies derived from gradient correlations of different local clients.

\begin{figure}[!t]
\centering
\includegraphics[width=0.6\textwidth]{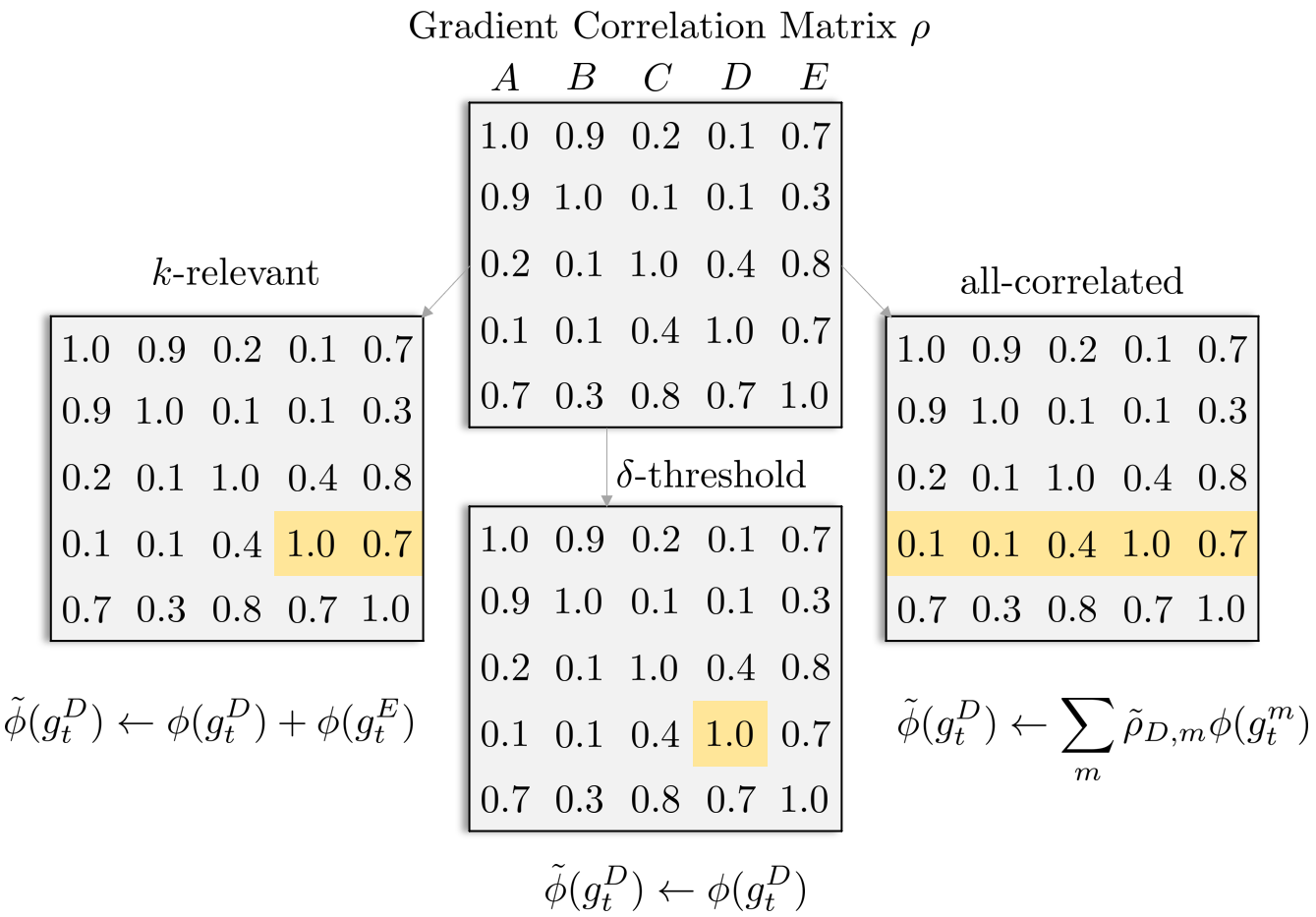}
\caption{A toy example of our personalized aggregation strategies on client D. We set $k=2$ in the $k$-relevant strategy and $\delta>0.8$ in the $\delta$-threshold strategy. Besides, $\tilde{\rho}$ denotes the softmax version of $\rho$.}
\label{strategies}
\end{figure}

Based on the received and compressed gradient information of the local clients, we measure their similarity using the Pearson correlation coefficient. The obtained gradient correlation coefficient matrix $\boldsymbol{\rho}$ is denoted as
\begin{equation}\boldsymbol{\rho}=
\begin{bmatrix}
\rho_{1,1} & \rho_{1,2} & \cdots & \rho_{1,|\mathcal{M}_t|}\\
\rho_{2,1} & \rho_{2,2} & \cdots & \rho_{2,|\mathcal{M}_t|} \\
\vdots & \vdots & \ddots & \vdots \\
\rho_{|\mathcal{M}_t|,1} & \rho_{|\mathcal{M}_t|,2} & \cdots & \rho_{|\mathcal{M}_t|,|\mathcal{M}_t|}
\end{bmatrix}.
\end{equation}
For any given client $m$ and $s$ and their corresponding gradients $\phi(g_t^m)$ and $\phi(g_t^s)$, their correlation is computed as
\begin{equation}
\rho_{m,s} = \frac{cov(\phi(g_t^m), \phi(g_t^s))}{\sigma_{\phi(g_t^m)} \sigma_{\phi(g_t^s)}},
\end{equation} 
where $cov(\cdot)$ denotes the covariance, and $\sigma$ is the standard deviation.

Intuitively, $\boldsymbol{\rho}$ measures the distances of gradients among different local clients, it also reflects the similarities of wireless traffic patterns, i.e., the spatial dependencies of local clients.
To be blunt, if a client is highly correlated with the others, then its contribution to the final global model should be enhanced, otherwise, it should be weakened. This is because if a client has very small correlations with other clients, then it belongs to the category outliers, whose gradient information may considerably affect the global optimization.
Thus based on this principle and the correlation matrix $\boldsymbol{\rho}$, we propose a family of personalized gradient aggregation strategies, which can be expressed as follows
\begin{equation}
\tilde{\phi}(g_t^m) = f_{agg}\lbrace \phi(g^{s}_t)| s \in \mathcal{M}_t \rbrace,
\end{equation}
where $f_{agg}(\cdot)$ represents different functions to assist a client $m$ in incorporating gradient information from others.
Specifically, we design three strategies on the basis of the gradient correlation matrix, i.e., $k$-relevant, $\delta$-threshold, and all-correlated. A toy example of these three personalized aggregation strategies can be found in Fig. \ref{strategies} and the details are given in the following paragraphs.

\subsubsection{$k$-relevant strategy}
In this strategy, we select the gradient information of $k$ clients in total, based on the strength of the correlation. As shown in Fig. \ref{strategies}, when $k=2$, the most correlated two clients for client D are E and itself. Thus client D will incorporate the gradient information of client E. Generally speaking for any client $m$, this strategy can be described as 
\begin{equation}\label{s1}
\tilde{\phi}(g_t^m) = \sum_{s \in \mathcal{M}_t} \mathbb{I}(\rho_{m,s}\geqslant \texttt{maxk}(\rho_{m}))\cdot \phi(g_t^s),
\end{equation}
where $\texttt{maxk}(\cdot)$ calculates the $k$-th largest value of a vector and the indicator function $\mathbb{I}(\cdot)=1$ when $\rho_{s,m} \geqslant \delta$ and 0 otherwise.

\begin{algorithm}[!t]
\DontPrintSemicolon
\KwInput{Local datasets $\{\mathcal{D}^m\}_{m=1}^M$, number of communication rounds $T$, number of local updates $\tau$, learning rates $\eta$ and $\epsilon$, initial global model $w_{0}$ and aggregated gradients $g_0$, error feedback $e_{0}$, and gradient tracking $h_{0}$
}
\KwOutput{Global model $w$}
\While{$t < T$}
{
	Sample a subset clients $\mathcal{M}_t$ \\
	\Comment {Client side}
  \For {each $m \in  \mathcal{M}_t$ in parallel} 
  { 
  		Initialize local model $w^m_{t,0}=w_t$ \\
  		\For {$c=0,1,\cdots,\tau-1$}
  		{
	  		Sample a batch data samples $\mathcal{B}_{t,c}^m$ \\
	  		$g_{t, c}^m = e_t^m + \nabla f_m(w_{t,c}^m;\mathcal{B}_{t,c}^m) - h_t^m$ \\
	  		$w_{t,c+1}^m = w_{t,c}^m - \epsilon g_{t, c}^m$ \\
  		}
  		Calculate the accumulated gradient by $g_t^m = (w_t^m - w_{t,\tau}^m)/\epsilon$ and send $\phi (g_t^m)$ to the central server \\
  		Update error feedback: $e_{t+1}^m = e_{t}^m + g_t^m - \phi (g_t^m)$ \\
  		Update gradient tracking: $h_{t+1}^m = h_{t}^m + \frac{1}{\tau}(\phi(g_t^m) - g_t)$ \\
  		Update global model locally: $w_{t+1}=w_t - \eta g_{t}$
  		
  }
  \Comment {Server side}
  Choose a strategy $f_{agg}(\cdot)$ from (\ref{s1}), (\ref{s2}), or (\ref{s3}) \\
  \For {each $m \in \mathcal{M}_t$}
  {
  		$\tilde{\phi}(g_t^m) = f_{agg}\lbrace \phi(g^{s}_t)| s \in \mathcal{M}_t \rbrace$
  }
  Server aggregates local gradients by $g_{t} = \frac{1}{|\mathcal{M}_t|}\sum_{m\in \mathcal{M}_t}\tilde{\phi}(g_t^m)$ and broadcast it to each local client \\
}
\caption{Gradient compression- and correlation-driven FL for wireless traffic prediction}
\label{fedgcc_alg}
\end{algorithm}

\subsubsection{$\delta$-threshold strategy}
In this strategy for arbitrary client $m$, we select the clients that have correlation coefficients higher than threshold $\delta$. Take also the client D in Fig. \ref{strategies} as an example. The personalization will not change its gradient information when we set $\delta=0.8$, since all its correlations with the other clients are smaller than this threshold. As a result, its personalized version of the gradient information remains the same as before. We describe this strategy as
\begin{equation}\label{s2}
\tilde{\phi}(g_t^m) = \sum_{s\in \mathcal{M}_t} \mathbb{I}(\rho_{m,s}\geqslant \delta) \cdot \phi(g_t^s),
\end{equation}
where $\mathbb{I}(\cdot)$ is also an indicator function that returns 1 when $\rho_{m,s}\geqslant \delta$ and 0 otherwise.

\subsubsection{All-correlated strategy}
In this strategy, each client aggregates all the other client's gradient information using a weighted average and the weights are derived from the correlation matrix $\boldsymbol{\rho}$. Take client D in Fig. \ref{strategies} for example. Client D will update its gradient and take all other clients' information for consideration but weigh the importance of clients C and D since they have larger correlations with client D.
For any client $m$, the general update rule for this strategy is
\begin{equation}\label{s3}
\tilde{\phi}(g_t^m) = \sum_{s \in \mathcal{M}_t}\tilde{\rho}_{m,s} \phi(g_t^s),
\end{equation}
with $\tilde{\rho}_{m,s}$ obtained by
\begin{equation}
\tilde{\rho}_{m,s} = \frac{e^{\rho_{m,s}}}{\sum_{v\in \mathcal{M}_t} e^{\rho_{m,v}} }.
\end{equation}

Having obtained the personalized gradient information for each client, the final global optimization at the server can be performed as follows
\begin{equation}
w_{t+1} = w_t - \frac{\eta}{|\mathcal{M}_t|}\sum_{m\in \mathcal{M}_t} \tilde{\phi}(g_t^m).
\end{equation}

The whole updates and procedures are summarized in Algorithm 1.

\section{Experiment Results}
In this section, we perform experiments on two real-world wireless traffic datasets. We first describe and analyze the two datasets briefly. Then, we provide a detailed explanation of parameter settings, evaluation metrics, and the corresponding baselines. Lastly, we present the prediction performance of our proposed algorithm and its comparisons with state-of-the-art models.

\begin{table}
\renewcommand\arraystretch{1.3}
\centering
\caption{Statistics of the two real-world cellular traffic.}
\label{tab:statistics}
\resizebox{0.4\textwidth}{!}{%
\begin{tabular}{@{}lcc@{}}
\toprule
                        & Milan                 & Trentino                \\ \midrule
Time span               & \multicolumn{2}{c}{From Nov. 1, 2013 to Jan. 1, 2014} \\ 
Time interval               & \multicolumn{2}{c}{10 minutes} \\ \midrule
\# of BSs              & $4,222$                  & $1,466$                 \\
\# of SCUs         & $88$                     & $223$                   \\ 
Average traffic volume  & $7,250$ MB                & $608$ MB                \\ 
Coefficient of variation & $1.1377$                 & $2.3410$                \\ \bottomrule
\end{tabular}%
}
\end{table}

\begin{figure}[!t]
\centering
\subfloat[]
{\includegraphics[width=0.35\textwidth]{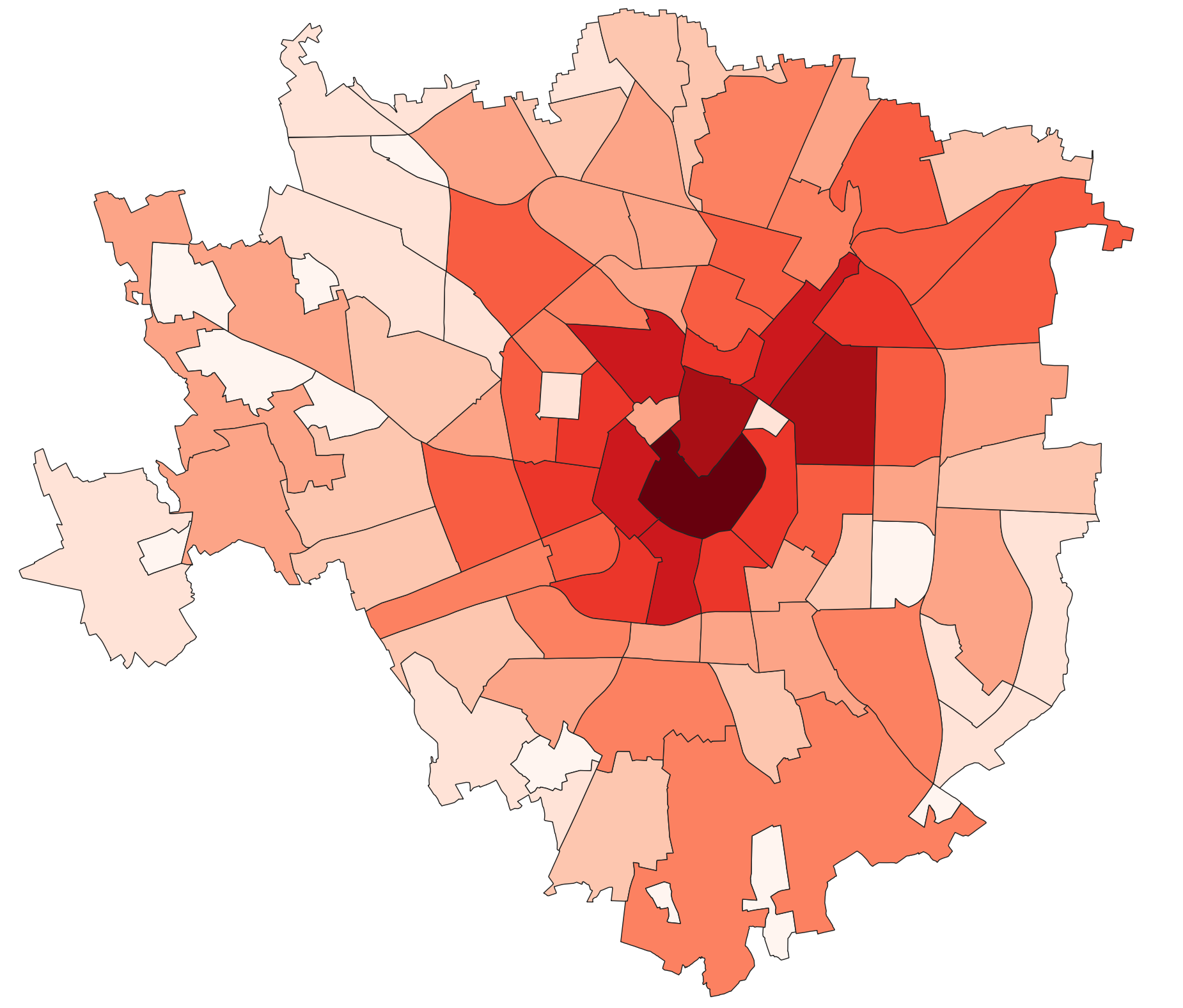}
\label{milan_traffic}
}

\subfloat[]
{\includegraphics[width=0.35\textwidth]{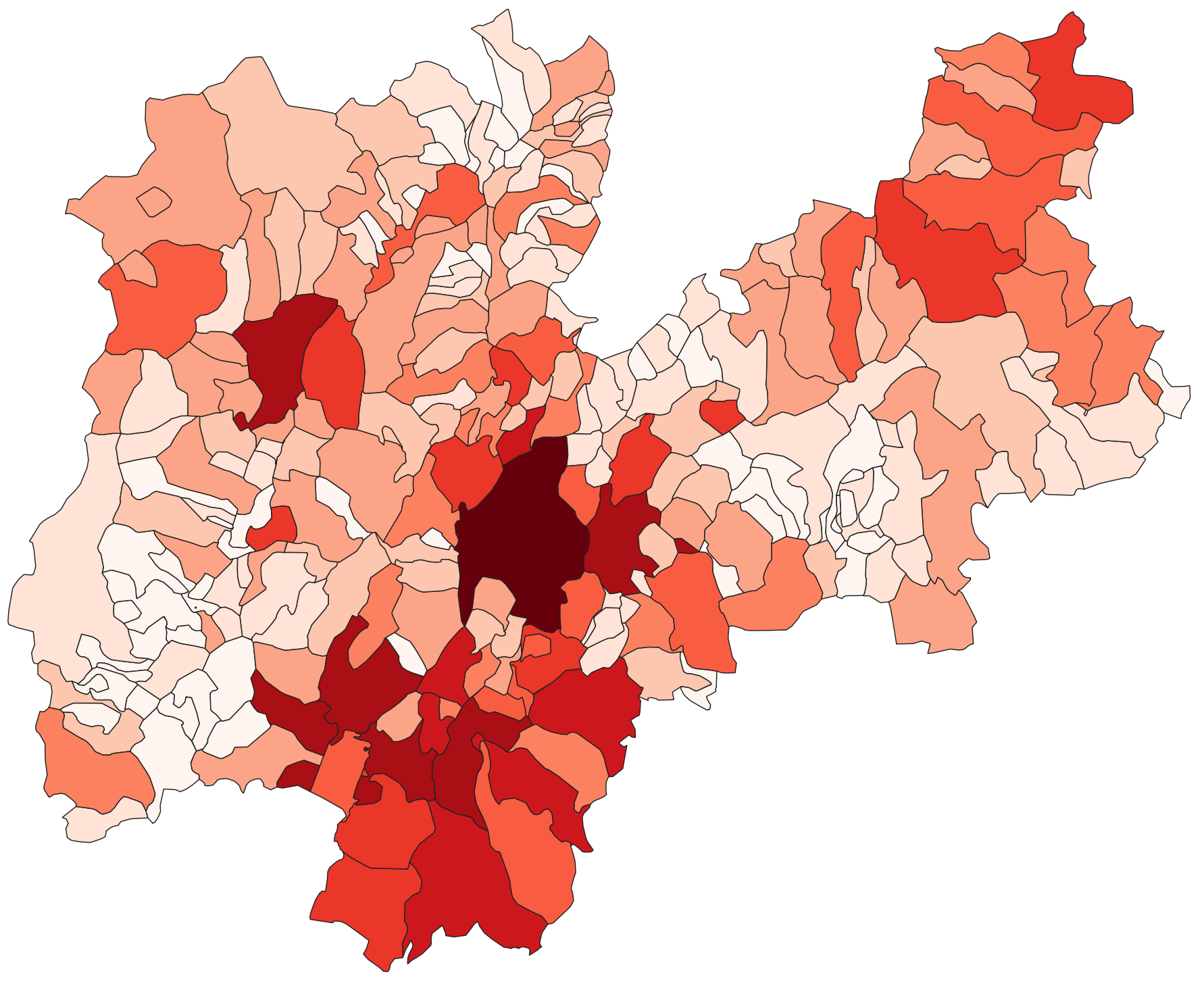}
\label{trentino_traffic}
}
\caption{Average traffic distribution of the two real-world datasets. (a) Milan; (b) Trentino. The darker the color, the larger the traffic volume.}
\label{traffic_distribution}
\end{figure}

\subsection{Datasets Description}
The datasets adopted in this study are Call Detail Records (CDRs) from the city of Milan, Italy and the province of Trentino, Italy, collected every 10 minutes over a two-month time span and released by Telecom Italia \cite{Barlacchi2015}. In the following description, we denote them as Milan and Trentino, respectively.
The raw CDRs are geo-referenced, anonymized and aggregated Internet traffic data based on the location of BSs. Specifically, a CDR record is logged if a user transfers more than 5~MB of data or spends more than 15 minutes online. For the convenience of understanding, we use MB as the unit of the data.
Table \ref{tab:statistics} shows the detailed statistics of these two datasets and Fig. \ref{traffic_distribution} gives the spatial distribution of the averaged wireless traffic.
Note that though the average traffic volume of the Milan is 7,250 MB, which is much higher than that of the Trentino, from the perspective of the coefficient of variation, the Trentino dataset is larger than the Milan dataset, indicating high heterogeneity among different SCUs.

\begin{table*}[!t]
\centering
\caption{Prediction performances of different algorithms on the Milan and Trentino datasets. The best prediction results are marked as bold for easy comparison. $\Delta C$ shows the total transferred data bytes between local clients and the central server measured in MB. }
\renewcommand\arraystretch{1.3}
\label{tab:pred}
\resizebox{0.95\textwidth}{!}{%
\begin{tabular}{llccccccc}
\toprule
\multirow{2}{*}{Method} & \multirow{2}{*}{Notes} & \multicolumn{3}{c}{Milan} & \multicolumn{3}{c}{Trentino} & \multirow{2}{*}{$\Delta C$ (MB)}\\ \cmidrule{3-8}
 &              & RMSE   & MAE    & $R^2$ Score & RMSE   & MAE    & $R^2$ Score \\ \midrule
FedAvg \cite{McMahan2017} & -        & $0.1401$ & $0.0965$ & $0.9371$   & $1.0504$ & $0.5834$ & $0.7871$ & $126.27/322.68$  
\\ \midrule
\multirow{3}{*}{FedProx \cite{Li2020}} & $\mu=0.01$  & $0.1398$ & $0.0960$ & $0.9373$   & $1.0057$ & $0.5581$ & $0.8129$ & \multirow{3}{*}{\begin{tabular}[c]{@{}l@{}}$126.27/322.68$\end{tabular}}  \\
                         & $\mu=0.1$   & $0.1400$ & $0.0963$ & $0.9373$   & $1.0402$ & $0.5774$ & $0.7936$  \\
                         & $\mu=1$     & $0.1339$ & $0.0869$ & $0.9446$   & $1.1615$ & $0.6475$ & $0.7114$  \\ \midrule
FedAtt \cite{Ji2019} & - & $0.1357$ & $0.0898$ & $0.9431$   & $0.9194$ & $0.5269$ & $0.8637$ & $126.27/322.68$  \\ \midrule
\multirow{3}{*}{FedDA \cite{zhang2021fedda}}   & $\varphi=1$  & $0.1339$ & $0.0816$ & $0.9466$   & $0.7391$ & $0.3933$ & $0.9264$ & $126.30/322.74 $ \\
& $\varphi=10$  & $0.1308$ & $0.0795$ & $0.9493$   & $0.7823$ & $0.4188$ & $0.9143$ & $126.55/323.39$  \\
                         & $\varphi=100$ & $0.1301$ & $0.0790$ & $0.9493$   & $0.7711$ & $0.3918$ & $0.9217$  & $129.06/329.81$ \\ \midrule
FedCOMGATE \cite{Haddadpour2021} & - & $0.1438$ & $0.1027$ & $0.9317$   & $0.7427$ & $0.3849$ & $0.9273$  & $14.373/38.449$  \\ \midrule
                         
\multirow{3}{*}{$\mathsf{Proposed}$} & $k$-relevant & $\bf{0.1299}$ & $\bf{0.0788}$ & $\bf{0.9501}$   & $\bf{0.6935}$ & $\bf{0.3621}$ & $0.9431$ & \multirow{3}{*}{\begin{tabular}[c]{@{}l@{}}$\bf{3.1494}/\bf{7.5843}$\end{tabular}}  \\
  & $\delta$-threshold& $0.1301$ & $0.0797$ & $0.9486$   & $0.6943$ & $0.3638$ & $0.9423$   \\
   & all-correlated & $0.1300$ & $0.0795$ & $0.9483$   & $0.7048$ & $0.3805$ & $\bf{0.9499}$   \\ \bottomrule
\end{tabular}%
}
\end{table*}

\subsection{Baseline Algorithms and Evaluation Metrics}
We compare our method with the following state-of-the-art algorithms.
\begin{itemize}
\item FedAvg \cite{McMahan2017}. This is the seminal work on federated learning from decentralized data. The clients perform local SGD multiple times and send the obtained gradients or updated parameters to the server, where the aggregation is performed to yield the final model.
\item FedProx \cite{Li2020}. It was proposed to stabilize the training of FedAvg on Non-IID data by adding a proximal term on the objective function of local clients.
\item FedAtt \cite{Ji2019}. By this method, the server performs a weighted aggregation instead of a simple average, and the weights are obtained based on the Euclidean distance between the global model and the local model.
\item FedDA \cite{zhang2021fedda}. This method represents a dual-attention based federated optimization algorithm. When updating the global model, the server considers the Euclidean distances between local models and the global model and the quasi-global model.
\item FedCOMGATE \cite{Haddadpour2021}. This is the state-of-the-art FL algorithm with gradient compression techniques. It provides theoretical guarantees on the convergence of FL when introducing gradient compression. FedCOMGATE is designed for general machine learning problems, not for wireless traffic prediction.
\end{itemize}

To quantitatively evaluate the performances of different wireless traffic prediction algorithms, we adopt three widely used metrics in the following experiments, i.e., Root Mean Square Error (RMSE), Mean Absolute Error (MAE), and R Squared ($R^2$) score. 
RMSE and MAE denote the difference between predictions and ground truth. Smaller values denote better prediction results. R squared score provides a measure of how well predictions are replicated by the model, based on the proportion of total variation of predictions explained by the model. The higher the values, the better the prediction results. Additionally, we also compare the amount of transferred gradients ($\Delta C$) between local clients and the central server of different algorithms.

\subsection{Experiment Settings}
The experiment results are obtained with the following settings. 
We first utilize a sliding window scheme to construct training and test datasets. Specifically, data from the first seven weeks are used to train the model, and all remaining data are used to test prediction performance. The window size $p$ is set to 6, that is, we use one-hour Internet traffic as feature input to predict the traffic volume of the next time slot. 
To accelerate training, all data samples are standardized by subtracting the mean and dividing by the standard deviation for each SCU.
Considering the necessity to minimize the extra computing and communication burdens of model training on wireless networks, a relatively lightweight prediction model is preferred for edge clients. Thus a three-layer MLP architecture is designed with hidden neurons of 128, 128, and 1, respectively.

As for the setting of hyper-parameters involved in FL, we follow the instructions in previous works \cite{McMahan2017,Haddadpour2021,zhang2021fedda}.
Specifically, we use stochastic gradient descent (SGD) as the optimizer to train our model and train it with 200 communication rounds in total.
For the learning rate of local updates, i.e., $\epsilon$, we start with 0.1 and decay it by 10 using a multi-step learning rate scheduler with milestones of 100 and 150. The learning rate at the server, $\eta$, is set to 1.0.
Before sending the gradients to the server for aggregation, we update the local model via 5 steps of SGD with a batch size of 20 and compression ratio of 0.01. In our $k$-relevant and $\delta$-threshold strategies, we set $k=4$ and $\delta=0.5$ unless otherwise specified.
Note that in order to make fair performance comparisons, all prediction models, including our model and all the baseline models, use exactly the same network architecture and parameter settings. We implement all algorithms using PyTorch \cite{Paszke2019} library and run the script on a desktop running a CentOS system.

\subsection{Overall Prediction Performance}\label{opp}

Table \ref{tab:pred} reports the achieved prediction performances of different algorithms. 
It is worth noting that in $\texttt{FedProx}$, there is a parameter $\mu$ that balances the significance of the proximal term. Similarly, $\texttt{FedDA}$ holds a parameter $\varphi$ to control how many augmented data samples should be transferred to the server for quasi-global model training. We consider different choices of these parameters and report the corresponding results in Table \ref{tab:pred}.

From this table, it is evident that our proposed algorithm achieves predominantly better prediction results than all the baselines. More specifically, different algorithms perform a kind of similar on the Milan dataset, but they vary a lot on the Trentino dataset. Take the MAE metric on the Milan dataset as an example, the best prediction result is 0.0788, obtained by our method with the $k$-relevant aggregation strategy. $\texttt{FedDA}$ achieves the second best MAE result, that is, 0.0790, when transferring all the augmented data samples ($\varphi=100$) to the server side. This indicates our method gains approximately the same performance as the best baseline algorithm. For the $R^2$ score achieved on the Milan dataset, our best result is 0.9501, which is comparable to the second-best one (0.9493) achieved by $\texttt{FedDA}$.

\begin{figure*}[!t]
\centering
\subfloat[]
{\includegraphics[width=0.45\textwidth]{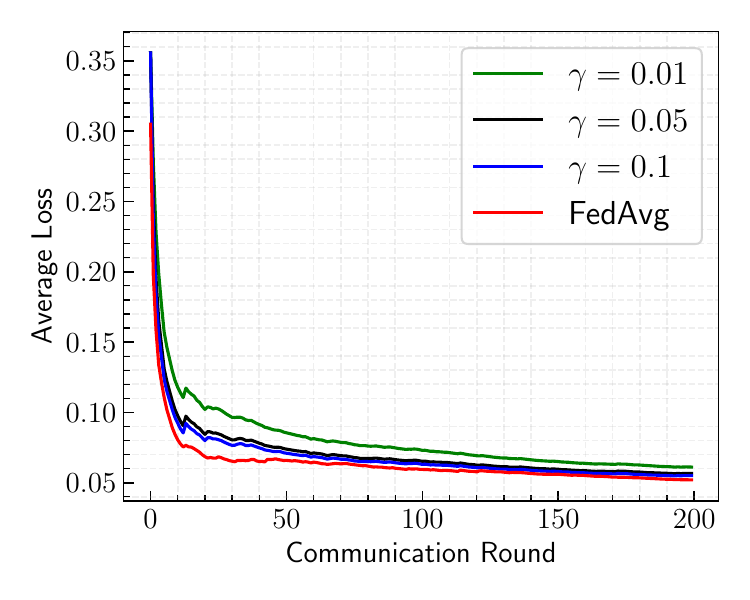}
\label{milan_loss}
}
\subfloat[]
{\includegraphics[width=0.45\textwidth]{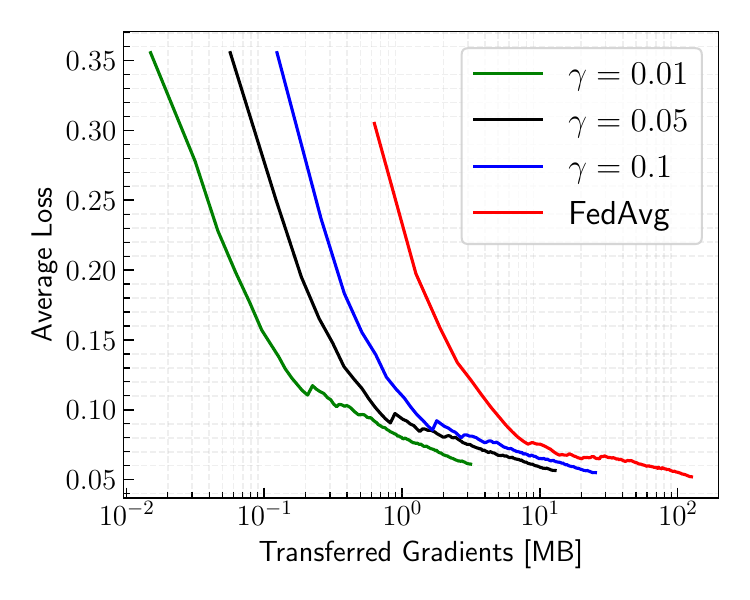}
\label{milan_grad}
}

\subfloat[]
{\includegraphics[width=0.45\textwidth]{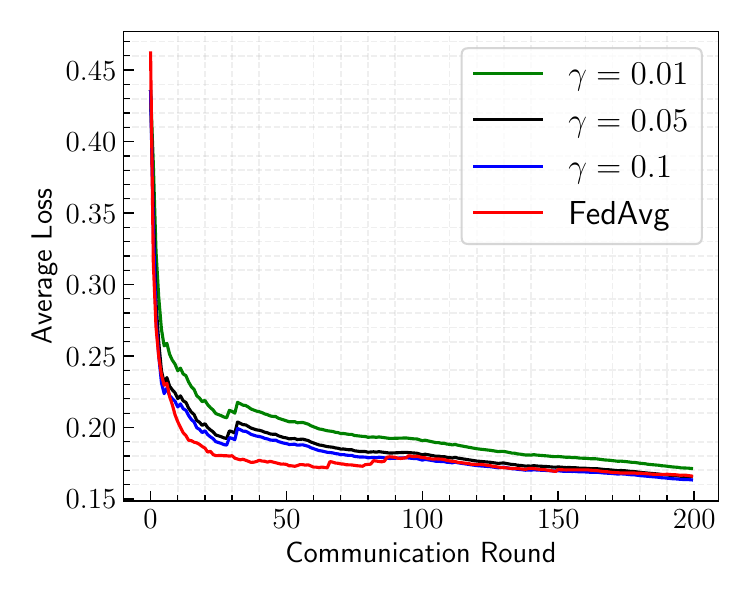}
\label{trentino_loss}
}
\subfloat[]
{\includegraphics[width=0.45\textwidth]{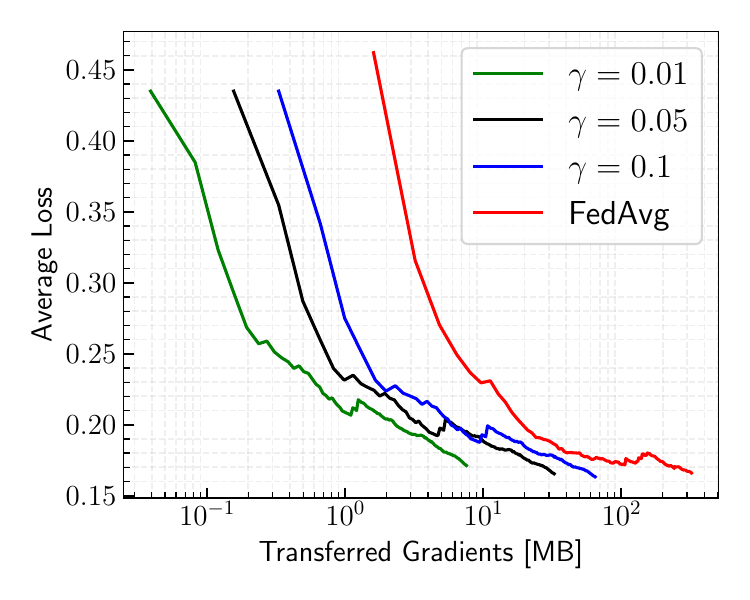}
\label{trentino_grad}
}
\caption{Convergence and communications on the two datasets: (a) Loss versus communication rounds on the Milan dataset; (b) Loss versus transferred gradient size on the Milan dataset; (c) Loss versus communication rounds on the Trentino dataset; (d) Loss versus transferred gradient size on the Trentino dataset.}
\label{loss_vs_epoch}
\end{figure*}

Meanwhile, on the Trentino dataset, our algorithm achieves noticeable performance improvements over baselines. For instance, the best MAE result of our algorithm is 0.3621, obtained under the scenario of $k$-relevant aggregation strategy. This indicates there is at least a 5.9\% MAE improvement over the best baseline's MAE result, which is 0.3849 obtained by $\texttt{FedCOMGATE}$ algorithm. 
Likewise, our algorithm improves the $R^2$ score from 0.9273, achieved by $\texttt{FedCOMGATE}$ method, to 0.9499, obtained by the all-correlated aggregation strategy.

Besides the accuracy performance, we can also observe the communication performances of different algorithms in Table \ref{tab:pred}. To be more concrete, $\texttt{FedProx}$ and $\texttt{FedAtt}$ have the same communications as $\texttt{FedAvg}$ since all of them need only transfer the gradients of each SCU to the server and no extra communications are required.
Though $\texttt{FedDA}$ achieves the best prediction performance among all baselines, it requires slightly more communications to transfer part of the augmented data samples to the server for quasi-global model training.
Our algorithm only needs to transfer the compressed gradient and the corresponding gradient indices, thus it is more communication-efficient when compared to $\texttt{FedDA}$. 
Moreover, our algorithm has a lower communication complexity than the classical federated compression algorithm $\texttt{FedCOMGATE}$, which also needs to transfer a control variable for gradient update. Though $\texttt{FedCOMGATE}$ achieves the lowest communication complexity among all baselines, its prediction performance is still inferior to ours since the personalized aggregation strategies enable our algorithm to capture the non-uniform spatial correlations of different SCUs. Consequently, our proposed algorithm achieves better prediction performance than $\texttt{FedCOMGATE}$. 

\begin{figure*}[htp]
\centering
\subfloat[]
{\includegraphics[width=0.63\textwidth]{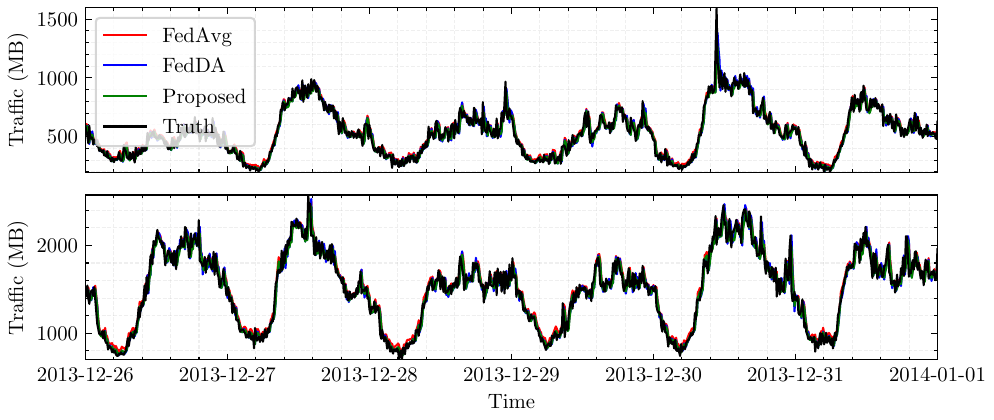}
\label{milan_pred}
}
\subfloat[]
{\includegraphics[width=0.33\textwidth]{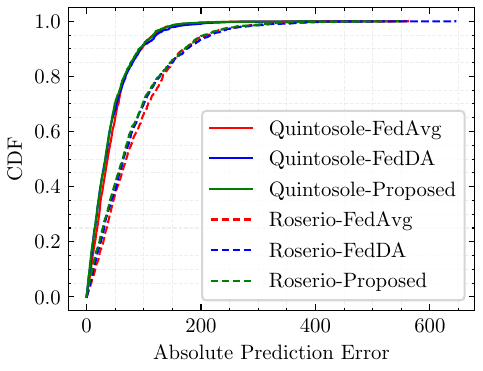}
\label{milan_cdf}
}

\subfloat[]
{\includegraphics[width=0.63\textwidth]{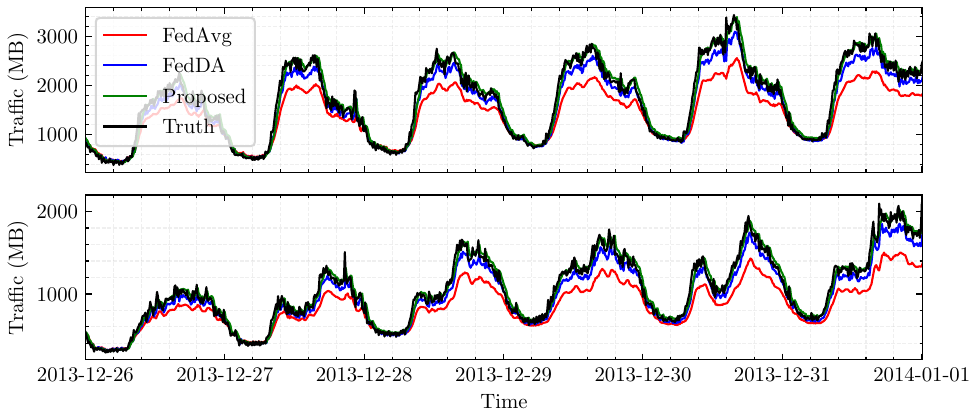}
\label{trentino_pred}
}
\subfloat[]
{\includegraphics[width=0.33\textwidth]{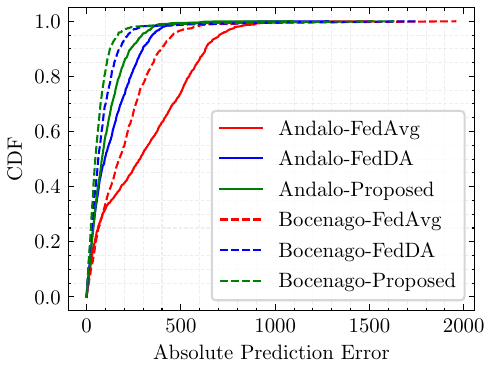}
\label{trentino_cdf}
}
\caption{Comparisons between the predicted and ground truth values: (a) Quintosole (upper) and Roserio (lower); (b) Prediction error analysis for Quntosole and Roserio; (c) Andalo (upper) and Bocenago (lower); (d) Prediction error analysis for Andalo and Bocenago.}
\label{pred_vs_truth}
\end{figure*}

\subsection{Convergence and Communication Efficiency Analysis}
In this subsection, we analyze the convergence and communication efficiency of our algorithm in terms of training loss and transferred gradient size, respectively. The obtained results are displayed in Fig. \ref{loss_vs_epoch}. Note that we only compare our algorithm with $\texttt{FedAvg}$ for simplicity. As the compression ratio influences the performance of our algorithm the most, we report the communication efficiencies under different compression ratios, that is, $\gamma \in \{0.01, 0.05, 0.1\}$. 

We can see from Fig. \ref{milan_loss} and Fig. \ref{trentino_loss} that our algorithm achieves a larger loss than $\texttt{FedAvg}$ at the initial phase of model training (e.g., less than 50 rounds). This is because $\texttt{FedAvg}$ involves no gradient compression, thus it sends all the gradient information to the server for federated optimization, which results in faster convergence and a lower training loss. However, as the training continues, our algorithm gradually approaches $\texttt{FedAvg}$ and finally outperforms it since not only all gradient information will be sent to the server, but also the spatial correlations of different SCUs are captured by our personalized aggregation strategy.

As illustrated in Fig. \ref{milan_grad} and Fig. \ref{trentino_grad}, under certain training loss constraints, our algorithm can relieve a large portion of bandwidth when training models. Considering the Trentino dataset as an example. Our algorithm needs approximately 3 MB ($\gamma=0.01$), 4 MB ($\gamma=0.05$), 13 MB ($\gamma=0.1$) communications to reach the loss of less than 0.2. However, the bandwidths required for achieving the same performance by $\texttt{FedAvg}$ is over 30 MB. A similar phenomenon exists on the Milan dataset, and we omit the detailed explanations here.

\subsection{Prediction Performance on Different Single SCUs}
The above subsection demonstrates the superiority of our proposed algorithm from the perspectives of both prediction performance and communication complexity. But what is the prediction performance of our algorithm on a single SCU?  In this subsection, we go one step further and answer this question by reporting the comparisons between predictions and ground truth values on different SCUs randomly selected from the two datasets.

Without loss of generality, two SCUs per dataset are selected. The two selected SCUs for the Milan dataset are located in Quintosole and Roserio, respectively. In the meantime, for the Trentino dataset, the two selected SCUs are located in Andalo and Bocenago, respectively. The predictions versus ground truth comparisons for these four SCUs are illustrated in Fig. \ref{pred_vs_truth}. Note that these SCUs are randomly selected without considering their similarities or geographical locations. The Cumulative Distribution Functions (CDF) of absolute prediction errors for different SCUs are also included in Fig. \ref{pred_vs_truth}. Note that we only compare our algorithm with $\texttt{FedAvg}$ and $\texttt{FedDA}$, for the sake of clarity. Besides, all the results pertaining to the proposed method are obtained by the $k$-relevant aggregation strategy, unless otherwise specified.

\begin{figure*}[!t]
\centering
\subfloat[]
{\includegraphics[width=0.33\textwidth]{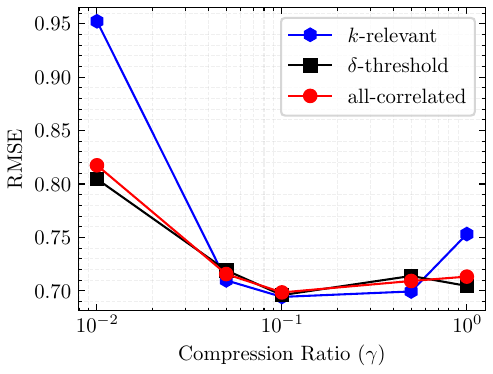}
\label{trentino_ratio}
}
\subfloat[]
{\includegraphics[width=0.33\textwidth]{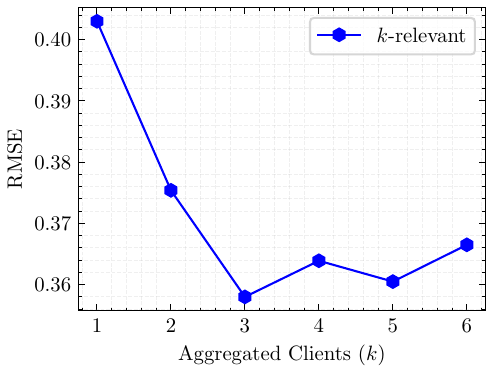}
\label{trentino_kbest}
}
\subfloat[]
{\includegraphics[width=0.33\textwidth]{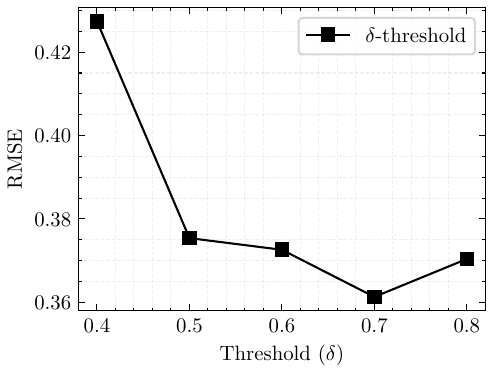}
\label{trentino_threshold}
}

\subfloat[]
{\includegraphics[width=0.33\textwidth]{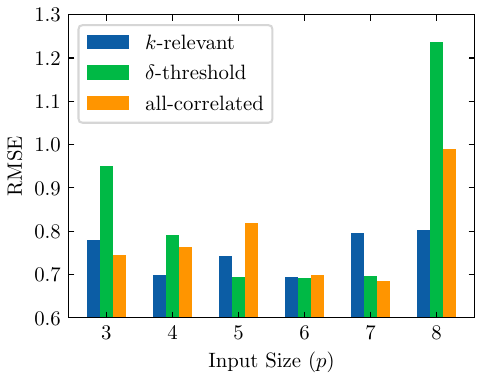}
\label{trentino_close}
}
\subfloat[]
{\includegraphics[width=0.33\textwidth]{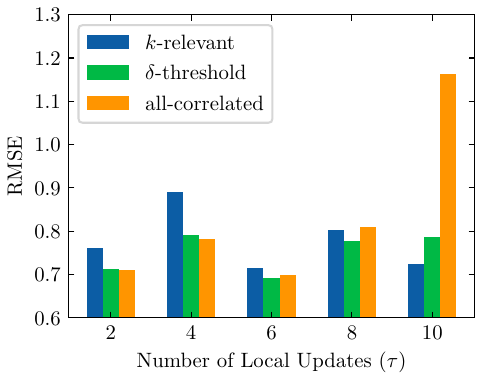}
\label{trentino_tau}
}
\subfloat[]
{\includegraphics[width=0.33\textwidth]{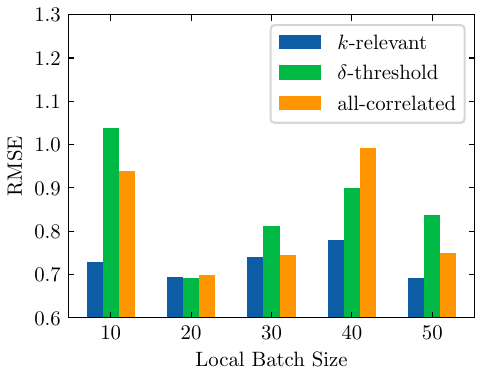}
\label{trentino_batch}
}
\caption{Parameter sensitivity on the Trento dataset: (a) RMSE versus compression ratio; (b) The change of $k$; (c) The change of threshold $\delta$; (d) RMSE versus input feature size; (e) RMSE versus the number of local updates; (f) RMSE performance when local batch size varies.}
\label{rmse_hyper}
\end{figure*}

For the two SCUs in the city of Milan, we can notice from Fig. \ref{milan_pred} that all three algorithms achieve excellent performances since the predictions and the ground truth values are matched well for both peak and non-peak hours. Technically speaking, our method achieves a slight improvement over $\texttt{FedAvg}$ and $\texttt{FedDA}$ algorithms, and it can be quantitatively reflected in Fig. \ref{milan_cdf}. Take Roserio as an example, the percentages of prediction errors that are less than 100 for $\texttt{FedAvg}$, $\texttt{FedDA}$, and our algorithm are 66\%, 70\%, and 72\%, respectively. For the two selected SCUs in the province of Trentino, we can observe from Fig. \ref{trentino_pred} and Fig. \ref{trentino_cdf} that our algorithm predicts the traffic values accurately, especially for the peak traffic values. The performance of $\texttt{FedAvg}$ is not satisfying due to its inability to cope with the data heterogeneity problem. $\texttt{FedDA}$ algorithm achieves better predictions than $\texttt{FedAvg}$, particularly for peak-hours traffic prediction, as it has prior knowledge of the data distribution and partially solves the data heterogeneity challenges. But $\texttt{FedDA}$ is still inferior to our algorithm since it failed to capture the personalized spatial correlations among different SCUs, which is of great importance for spatial temporal wireless traffic prediction. 
The results of Fig. \ref{pred_vs_truth} coincide with previous results of Table \ref{tab:pred}, thus again demonstrating the effectiveness of our algorithm in solving wireless traffic prediction problems.

\subsection{Parameter Sensitivity}
Several parameters in our algorithm will affect the final prediction performance, notably the compression ratio and input feature size. In this subsection, we report how different choices of these values influence the prediction results in Fig. \ref{rmse_hyper}. Note that we only show the RMSE results on the Trento dataset in Fig. \ref{rmse_hyper}, since MAE and $R^2$ score have very similar behaviors with the RMSE results on both datasets and their results are excluded in this figure to simplify the understanding of these parameters.

\subsubsection{Compression ratio}
The compression ratio affects prediction performance the most since it indicates how fast the server learns the complete gradient information of local clients. We consider different compression ratios like in Section \ref{opp} and report the achieved results of our three personalized aggregation strategies in Fig.~\ref{trentino_ratio}.
From this figure, we can observe that with the increase of compression ratio, for example, from 0.01 to 0.1, the RMSE performance improves gradually as more and more gradient information is transferred from the local client to the server. The server can learn the prediction model well when it receives enough gradient information.  However, keeping an increase of compression ratio does not necessarily bring more performance gains as the generalization ability of the prediction model does not improve. Experimentally, a compression ratio of 0.1 achieves relatively good performance.

\subsubsection{Aggregated clients $k$ and threshold $\delta$}
The results of different choices of $k$ and $\delta$ are shown in Fig.~\ref{trentino_kbest} and Fig.~\ref{trentino_threshold}, respectively, when setting compression ratio as 0.1. We can see from both figures that when too many clients are involved in model aggregation, performance is degraded. For example, when $k=6$ or $\delta=0.4$, the RMSE results are worse than $k=3$ and $\delta=0.7$, respectively. On the contrary, too few clients selected for model aggregation ($k=1$ or $\delta=0.8$) are not beneficial to performance as very limited information is used for model personalization. Thus there is a trade-off between performance and client involvement.

\subsubsection{Input feature size}
The input feature size also affects the predictions of our algorithm, and the results are displayed in Fig.~\ref{trentino_close}. We can understand from this figure that both too short and too-long input lengths are not suited for model training. Too short an input length may not properly reflect the relationships between the input and the target. In contrast, too long an input length may bring severe noise, making the model hard to learn. In our case, the prediction performance is relatively better when we use one hour's traffic ($n=6$) as input to predict the next time slot's traffic.

\subsubsection{Number of local updates $\tau$ and local batch size}
Fig.~\ref{trentino_tau} and Fig.~\ref{trentino_batch} show the results of RMSE regarding the number of local updates $\tau$ and local batch size. These two parameters control a trade-off between communication and computation.
A large $\tau$ or batch size results in more computations at local clients and thus reduces communications between local clients and the server. On the other hand, too much local computation may lower down the convergence rate or make the training not converge at all as each client goes too far towards its own gradient update direction. This could result in a poor-quality aggregated model distributed for training at local clients in the next ground. For example, in both figures of Fig.~\ref{trentino_tau} and Fig.~\ref{trentino_batch}, our model's prediction performance degrades when increasing $\tau$ from 6 to 8 or batch size from 20 to 40. 

From the above discussions, it is clear that there always exist trade-offs when setting parameters for the proposed FL framework and the relevant algorithms. In real-world tests and implementations, the optimal values of the above parameters can be obtained by applying grid search strategies.

\section{Conclusion}
In this work, we studied the distributed wireless traffic prediction at the edge for future communication networks and proposed a gradient compression- and correlation-driven FL algorithm to solve the formulated problem. Gradient compression, or more concretely, gradient sparsification techniques were introduced in our algorithm to release the communication overheads between local clients and the central server. In addition, we proposed using gradient correlation to reflect the traffic pattern similarities of geographically distributed local clients and derived three personalized gradient aggregation strategies, i.e., $k$-relevant, $\delta$-threshold, and all-correlated, for modeling spatial dependencies of different clients. We validated our proposed algorithm on two real-world datasets, and the results we obtained demonstrated the superiority of our algorithm over state-of-the-art methods in terms of both accuracy and communication for wireless traffic prediction.

\bibliographystyle{IEEEtran}
\bibliography{FedGCC}


 

\begin{IEEEbiography}[{\includegraphics[width=1in,height=1.25in,clip,keepaspectratio]{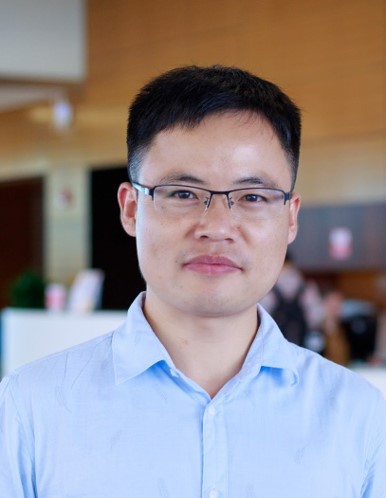}}]{Chuanting Zhang} (Senior Member, IEEE) received the Ph.D. degree in communication and information systems from Shandong University, Jinan, China, in 2019. 
He is currently an Associate Professor at the School of Software at Shandong University in Jinan, China. Before this, he was a Senior Research Associate at the University of Bristol, UK. Besides, he was a Post-Doctoral Fellow with the Computer, Electrical and Mathematical Science and Engineering Division, King Abdullah University of Science and Technology (KAUST), Thuwal, Saudi Arabia. He was the recipient of many awards, including the IEEE ICCT Young Scientist Award, the Excellent Doctoral Dissertation Award of Shandong Province, and IEEE SmartData Best Paper Award.
His current research interests include spatial-temporal data analysis, federated learning, AI for networks and networks for AI.
\end{IEEEbiography}

\begin{IEEEbiography}[{\includegraphics[width=1in,height=1.25in,clip,keepaspectratio]{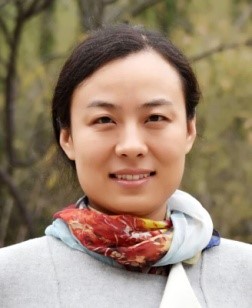}}]{Haixia Zhang} (Senior Member, IEEE) received the B.E. degree from the Department of Communication and Information Engineering, Guilin University of Electronic Technology, Guilin, China, in 2001, and the M.Eng. and Ph.D. degrees in communication and information systems from the School of Information Science and Engineering, Shandong University, Jinan, China, in 2004 and 2008, respectively. From 2006 to 2008, she was with the Institute for Circuit and Signal Processing, Munich University of Technology, Munich, Germany, as an Academic Assistant. From 2016 to 2017, she was a Visiting Professor with the University of Florida, Gainesville, FL, USA. She is currently a distinguished Professor with Shandong University, Jinan, China. Dr. Zhang is actively participating in many professional services. She is an editor of the IEEE Transactions on Wireless Communications, IEEE Wireless Communication Letters, and China Communications and serves/served as Symposium Chair, TPC Member, Session Chair, and Keynote Speaker at many conferences. Her research interests include wireless communication and networks, industrial Internet of Things, wireless resource management, and mobile edge computing.
\end{IEEEbiography}

\begin{IEEEbiography}[{\includegraphics[width=1in,height=1.25in,clip,keepaspectratio]{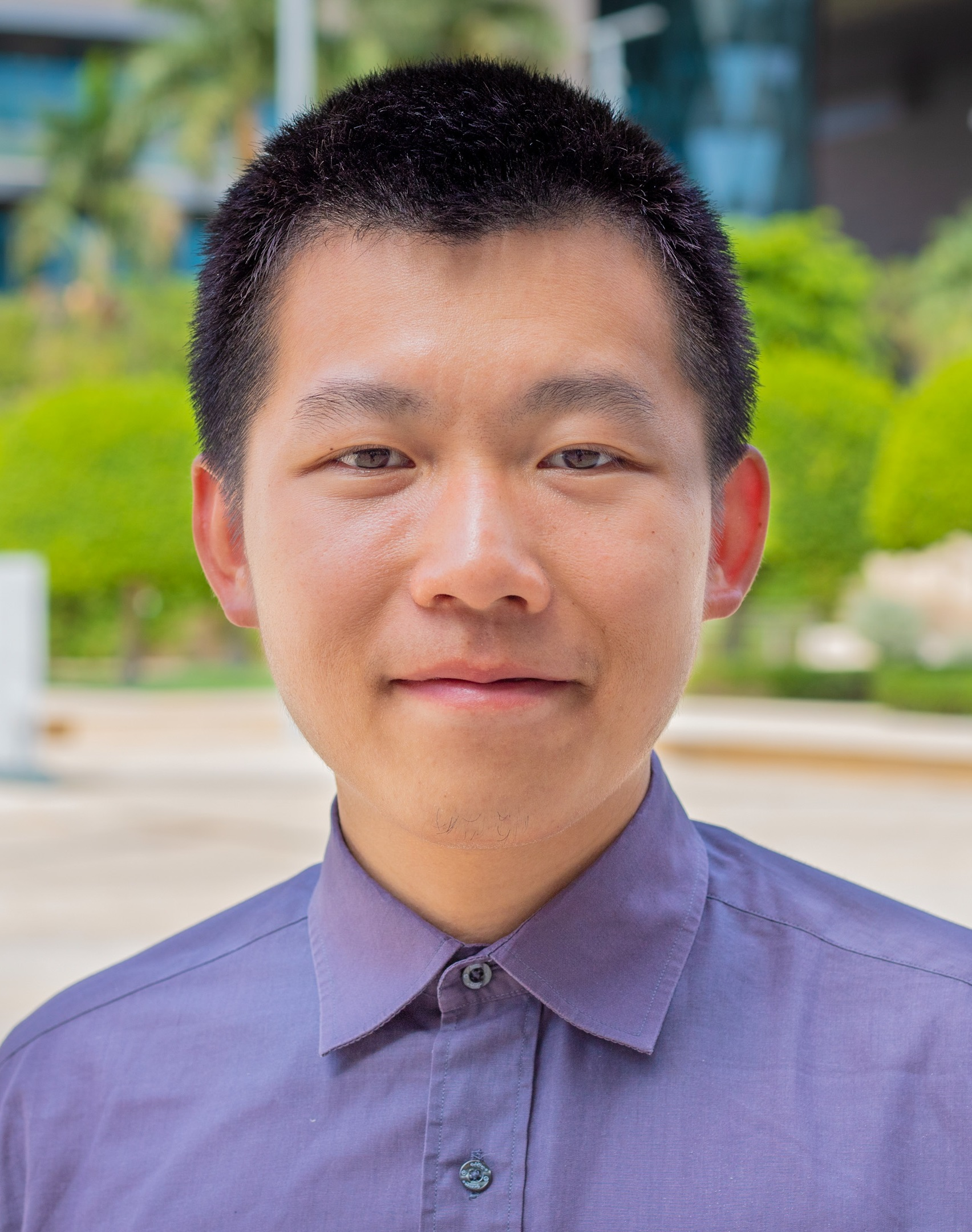}}]{Shuping Dang} (Member, IEEE) received B.Eng (Hons) in Electrical and Electronic Engineering from the University of Manchester (with first class honors) and B.Eng in Electrical Engineering and Automation from Beijing Jiaotong University in 2014 via a joint `2+2' dual-degree program. He also received D.Phil in Engineering Science from University of Oxford in 2018. Dr. Dang joined in the R$\&$D Center, Huanan Communication Co., Ltd. after graduating from University of Oxford and worked as a Postdoctoral Fellow with the Computer, Electrical and Mathematical Science and Engineering Division, King Abdullah University of Science and Technology (KAUST). He is currently a Lecturer with the Department of Electrical and Electronic Engineering, University of Bristol. The research interests of Dr. Dang include 6G communications, wireless communications, wireless security, and machine learning for communications. 
\end{IEEEbiography}

\begin{IEEEbiography}[{\includegraphics[width=1.1in,height=1.25in,clip,keepaspectratio]{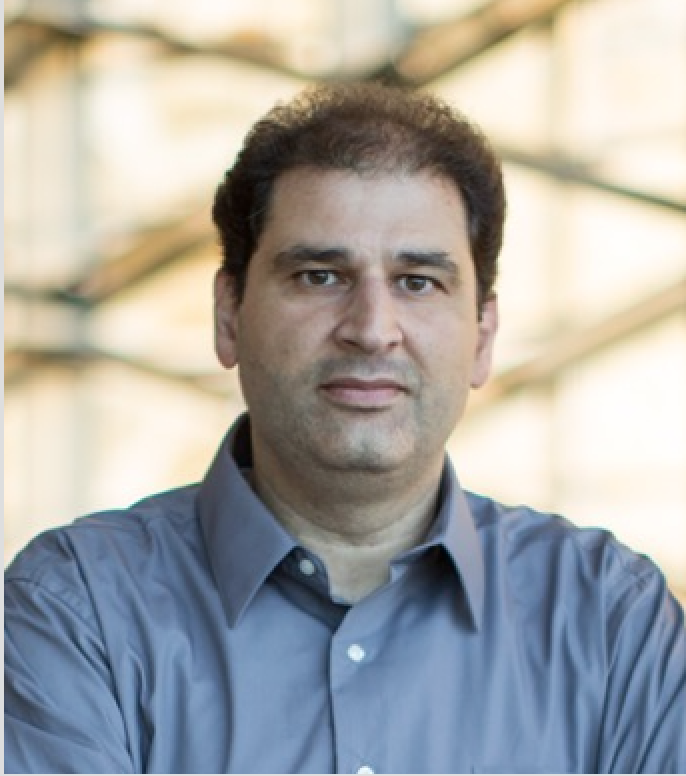}}]{Basem Shihada}(Senior Member, IEEE)  is an associate \& founding professor in the Computer, Electrical and Mathematical Sciences \& Engineering (CEMSE) Division at King Abdullah University of Science and Technology (KAUST). He obtained his PhD in Computer Science from University of Waterloo. In 2009, he was appointed as visiting faculty in the Department of Computer Science, Stanford University. In 2012, he was elevated to the rank of Senior Member of IEEE. His current research covers a range of topics in energy and resource allocation in wired and wireless networks, software defined networking, internet of things, data networks, network security, and cloud/fog computing. 
\end{IEEEbiography}

\begin{IEEEbiography}[{\includegraphics[width=5in,height=1.25in,clip,keepaspectratio]{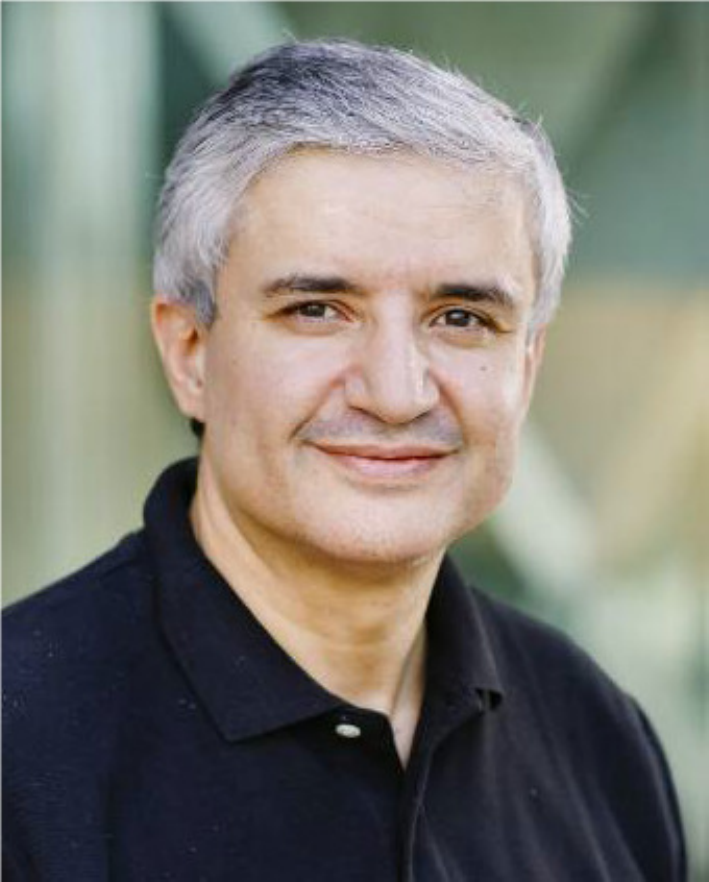}}]{Mohamed-Slim Alouini} (Fellow, IEEE) was born in Tunis, Tunisia. He received the Ph.D. degree in Electrical Engineering from the California Institute of Technology (Caltech), Pasadena, CA, USA, in 1998. He served as a faculty member in the University of Minnesota, Minneapolis, MN, USA, then in the Texas A\& M University at Qatar, Education City, Doha, Qatar before joining King Abdullah University of Science and Technology (KAUST), Thuwal, Makkah Province, Saudi Arabia as a Professor of Electrical Engineering in 2009. His current research interests include the modeling, design, and performance analysis of wireless communication systems.
\end{IEEEbiography}

\vfill

\end{document}